\documentclass[10pt,oneside,reqno]{amsart}
\title{El experimento de Cavendish}
\makeatletter
\g@addto@macro{\endabstract}{\@setabstract}
\newcommand{\authorfootnotes}{\renewcommand\thefootnote{\@fnsymbol\c@footnote}}%
\makeatother
\subjclass[2010]{00A30, 01A05, 58A05}
\usepackage{amsmath,amssymb,amsthm,graphicx,cancel,pifont,pb-diagram,mathpazo,epstopdf,wasysym,marvosym,mathrsfs,eso-pic,wallpaper,array,arydshln,multicol,etoolbox,morefloats,etex,mathtools,wrapfig}
\usepackage[activeacute,english,greek,spanish]{babel}
\usepackage[square,numbers]{natbib}
\usepackage[labelfont=bf]{caption}
\usepackage[font={small,it}]{caption}
\usepackage[belowskip=-10pt,aboveskip=6pt,labelsep=period]{caption}
\usepackage[LGR,T1]{fontenc}
\usepackage[10pt,GlyphNames,boldLipsian]{teubner}
\usepackage[titletoc,toc,page]{appendix}
\usepackage[scale=0.9]{ccicons}
\usepackage{fonttable}
\DeclareFontFamily{U} {MnSymbolC}{}
\DeclareFontShape{U}{MnSymbolC}{m}{n}{
    <-6>  MnSymbolC5
   <6-7>  MnSymbolC6
   <7-8>  MnSymbolC7
   <8-9>  MnSymbolC8
   <9-10> MnSymbolC9
  <10-12> MnSymbolC10
  <12->   MnSymbolC12}{}
\DeclareFontShape{U}{MnSymbolC}{b}{n}{
    <-6>  MnSymbolC-Bold5
   <6-7>  MnSymbolC-Bold6
   <7-8>  MnSymbolC-Bold7
   <8-9>  MnSymbolC-Bold8
   <9-10> MnSymbolC-Bold9
  <10-12> MnSymbolC-Bold10
  <12->   MnSymbolC-Bold12}{}
\DeclareSymbolFont{MnSyC} {U} {MnSymbolC}{m}{n}
\DeclareMathSymbol{\squaredots}{\mathrel}{MnSyC}{14}

\usepackage{tikz}

\usepackage{hyperref}
\hypersetup{bookmarksopen,bookmarksnumbered,colorlinks,linkcolor=blue,urlcolor=blue,citecolor=blue}
\newtheorem*{propo}{Proposición}

\begin{document}\maketitle
\begin{center}
\normalsize
\authorfootnotes
Jonathan Taborda Hern\'{a}ndez\footnote{\email{taborda50@gmail.com}}. \par \bigskip
\today
\end{center}
\begin{abstract}
En este art\'{i}culo, presentamos una descripci\'{o}n del \textit{apparatus} empleado por Henry Cavendish, el cual a su vez es en realidad un compendio de 17 complejos experimentos, para tratar de medir experimentalmente la constante de gravitación universal, planteada te\'{o}ricamente por el divino Sir Isaac Newton, en su monumental \textit{Principia Mathematica}.\\
Como la fuerza gravitacional es muy peque\~{n}a, los experimentos gravitacionales en el laboratorio son altamente susceptibles a perturbaciones extra\~{n}as. Medir la gravitaci\'{o}n en el laboratorio es entonces as\'{i} de problem\'{a}tico y en la actualidad tales dificultades persisten.
\end{abstract}
\selectlanguage{english}
\begin{abstract}
In this article, we present a description of the \textit{apparatus} employed by Henry Cavendish, which in turn is actually a compendium of 17 complex experiments, to try to experimentally measure the universal gravitation constant, theoretically posed by the divine Sir Isaac Newton, in his monumental \textit{Principia Mathematica}.\\
Since the gravitational force is very small, gravitational experiments in the laboratory are highly susceptible to strange disturbances. Measuring gravitation in the laboratory is then that problematic and today such difficulties persist.
\end{abstract}
\selectlanguage{spanish}
\tableofcontents
\section{Introducción: midiendo la fuerza gravitacional}
Comola fuerza gravitacional\index{Fuerza!gravitacional} es muy pequeña, los experimentos gravitacionales en el laboratorio son altamente susceptibles a perturbaciones extrañas. Medir la gravitación en el laboratorio es entonces así de problemático y en la actualidad tales dificultades persisten.\footnote{La precisión con la que la ley inverso-cuadrática puede ser establecida es sobre una parte en $10^4$, mientras que la ley inverso-cuadrática en electrostática es sobre una parte en $10^{16}$. Chen (\cite{1993}, p. 5) comentan lo siguiente: <<Una razón es la muy alta sensibilidad de las medidas eléctricas comparadas con las medidas mecánicas, la otra es el hecho de que los detectores eléctricos pueden ser completamente cerradoa en el interior de una celda de Faraday, mientras que no es posible construir una celda de Faraday gravitacional cerrada y así tener acceso a un detector mecánico>>.} En concordancia no tendrá que ser sorpresivo que un siglo después de la primera ed. del \textit{Principia}\index{Principia} de Newton\index{Newton} (1687) hasta la primera medida precisa de la gravitación fuera concebida y presentada. Newton por sí mismo se mostró excéptico sobre este tema y en su póstumamente publicado \textit{De mundi systemate} (1728) él tuvo que sugerir que, si dos esferas con un diámetro de 1 pie son colocadas a una distancia de $1/4$ cm una de la otra, ellas no entrarán en contacto por la fuerza de su atracción mutua en no menos de 1 mes.\footnote{Cf. Newton (\cite{Newton1728}, p. 27): <<Hujusmodi globi duo, quart\^{a} tant\`{u}m digiti parte ab invicem distantes, in spatis liberis, haud minori quam mensi unius intervallo, vi mutua attractionibus accedevent, ad invicem>>.} Además, él tuvo que notar que una montaña hemiesférica de 3 millas de altura y 6 millas de ancho, no logrará ser trazada por un péndulo a 2 min de su perpendicular primaria.\footnote{Ibid., p. 27: <<Sed nec montes toti suffecerint and sensibiles effectus: Ad redices montis hemispaerici alti tria milliaria $\&$ lati sex, pendulum vi montis attractum non deviabit scrupulis duobis primis a perpendiculo>>.}\\ La teoría newtoniana de la gravitación universal fue bien confirmada a distancias planetarias y, en un contexto significativo, esto fue fundado sobre observaciones astronómicas. En las prop. I-II del libro III del \textit{Principia}\index{Principia}, Newton\index{Newton} estableció que los planetas primarios son atraidos por una fuerza centripetal\index{Fuerza!centripetal} inverso-cuadrática dirigida \textit{quam proxime} hacia el centro del Sol y que los planetas circumsaturnianos y circumjuniales son atraidos por una fuerza centripetal inverso-cuadrática dirigida \textit{quam proxime} hacia el centro de Saturno y Júpiter, respectivamente Newton (\cite{1999}, p. 802).\footnote{Cf. Harper (\cite{Harper2011}) para una explicación detallada.} Esta conclusión fue garantizada por las observaciones astronómicas, que indicaban que los cuerpos celestes satisfacen la ley de áreas de Kepler\index{Kepler} (\textit{quam proxime}) y la ley armónica (exáctamente), y por la dependencia sistemática Newton\index{Newton} tuvo que establecer en las prop. I-IV del libro I del \textit{Principia}\index{Principia} entre, por un lado, la presencia de una ley inverso-cuadrática dirigida (\textit{quam proxime}) al centro del cuerpo atrayente y el cuerpo atraído describiendo la ley de áreas de Kepler (\textit{quam proxime}) y, por el otro lado, entre los tiempos periódicos variando como la potencia $3/2$ para el radio y la fuerza centripetal\index{Fuerza!centripetal} variando inversamente como el cuadrado del radio (cf. Newton \cite{1999}, pp. 444-451). Como una presentación esencial para la dependencia sistemática Newton\index{Newton} tuvo establecer en el libro I que ellas son rigurosamente deducidas de las leyes del movimiento, y así ser respaldado por ellas. En las prop. III-IV del libro III, Newton estableció que la Luna es atraída por una fuerza centripetal inverso-cuadrática hacia el centro de la Tierra. Dado que las observaciones astronómicas muestran que el movimiento de la Luna satisface la ley de áreas (\textit{quam proxime}) de Kepler\index{Kepler}, se sigue que~---dada la dependencia sistemática entre la ley de áreas y la presencia de una fuerza centripetal\index{Fuerza!centripetal}~---que la Luna es atraida por una fuerza cantripetal hacia el centro de la Tierra. Sin embargo, porque la Luna es un satélite solitario, Newton\index{Newton} no podría usar la ruta vía Corolario 6 para la Prop. IV del libro I~---como él tuvo que hacerlo en las proposiciones procedentes del libro III. Newton, sin embargo, mostró que, sobre la suposición de la ley inverso-cuadrática, la acceleración de la Luna en la región de la Tierra que viene a ser igual a la medida de Huygens para la acceleración terrestre, que fue una de las pocas indicaciones directas en soporte de la afirmación Newtoniana de que la gravitación era preservada en todas las direcciones bajo la superficie de la Tierra. Pese a que las Prop. I-IV del libro III principalmente involucraban <<deduciones de fenómenos>>, que fueron soportados por la dependencia sistemática como se estableció en el libro I (y, últimamente, por las leyes del movimiento), las Prop. V-VIII del libro III explícitamente contienen varios pasos deductivos que Newton\index{Newton} empleó en las \textit{regualae philosophandi}.\\ El salto más audaz, aunque teniendo algún soporte matemático en la Prop. LXIX del libro I, fue la generalización de Newton para las proposiciones que él tuvo que establecer para los cuerpos universalmente.\footnote{Newton como nosotros bien advertimos, por instancia, que la aplicación de la III ley para los cuerpos celestes y cuerpos universalmente fue una generalización inductiva. Cabe señalar que la teoría de la gravitación universal estuvo precedida por cuatro \textit{regula philosophandi}. Cf. Newton (\cite{1999}, p. 796).} Newton tuvo que hacer poco, excepto tomar la explicación para la acceleración en la superficie de la Tierra en la Prop. IV del libro III, sobre las fuerzas gravitacionales\index{Fuerza!gravitacional} entre cuerpos a pequeñas distancias.\footnote{Newton concibió un \textit{experimentum crucis}, que involucraba la medida de la superficie gravitacional, para decidir entre la teoría de la gravitación universal y la teoría vorticial. Newton estableció que, sobre la suposición de que la Tierra es una esfera achatada de densidad homogenea, la superficie gravitacional en el Ecuador resulta de la combinación de dos esferas, a saber la fuerza centrifugal\index{Fuerza!centripetal} (o en el Ecuador) y la fuerza gravitacional que surge de las fuerzas inverso-cuadráticas dirigidas hacia las partes individuales sobre una Tierra oblata (cf. Newton \cite{1999}, p. 830-831; Greenberg \cite{Greenberg1995}, pp. 1-14 para una detallada discusión). Por contraste, Christiaan Huygens, quien explicó la gravedad\index{Gravedad} en términos mecánicos, afirmó que las fuerzas centrifugales terrestres en el Ecuador solamente son suficientes para explicar las diferentes longitudes en segundos determinados por un péndulo. Por consecuencia, la variación de la superficie gravitacional con la latitud es más grande de acuerdo a la teoría newtoniana que con la de Huygens. Fue únicamente en el siglo XIX, sin embargo, que al asunto fue zanjado a favor de la gravitación universal.}\\ Esto entonces sirve para mostrar que la ley Newtoniana de la gravitación universal no podría fallar a pequeñas distancias y en este contexto el experimento de Cavendish\index{Experimento de Cavendish} y los experimentos relativos del siglo XIX jugaron un rol decisivo, pero ellos proporcionaron una evidencia fuertemente creciente y convergente para la universalidad de la teoría Newtonia de la gravitación. Desde esta perspectiva, emergen varios interrogantes histórico-sistemáticos interconectados:
\begin{enumerate}
\item[(1)] ?`Cómo fue la evidencia fuertemente creciente y convergente en favor de la teoría Newtoniana de la gravitación universal establecida?.\\
\item[(2)] ?`Qué hay sobre el desarrollo para los procedimientos experimentales seguidos y las características del apparatus experimental empleados?.\\
y finalmente,\\
\item[(3)] ?`Se tiene una unidad metodológica-experimental discernible en 100 años de investigaciones experimentales y, si es así, cómo se ve?.
\end{enumerate}
\par
Varios elementos de la ley de la gravitación universal de Newton\index{Newton} pueden estar sujetos a verificación empírica. Un experimento físico puede elegirse para verificar si la fuerza gravitacional\index{Fuerza!gravitacional} realmente varía inversamente proporcional al cuadrado de la distancia\footnote{Esto fue primero explícitamente verificado por Mackenzie (\cite{1895}, p. 334-339), quien mostró que no ocurren desviaciones en la ley inverso-cuadrática. Una investigación experimental reciente tuvo que confirmar dicho resultado. V. gr., en 2001 en la University of Washington la ley inverso-cuadrática de Newton\index{Newton} fue verificada para $218\mu m$ usando un anillo de metal, suspendido de un péndulo de torsión, y conteniendo 10 agujeros distribuidos equitativamente, (cf. Hoyle et al. \cite{Hoyle2001}).}, si la fuerza gravitacional es por lo tanto proporcional al producto de las masas involucradas, si la constitución de la masa hace una diferencia cuando las masas no están en el vacío, si la ley de gravitación universal\index{Ley de gravitación universal} es cierta a pequeñas distancias en lugar de distancias celestiales, y que valor para G es (y, inclusive, si G es realmente una constante). El trabajo experimental de los siglos XVIII y XIX, ha llevado a cabo la investigación de los dos últimos interrogantes.
\section{La génesis: el artículo de Cavendish sobre la densidad de la Tierra}
\begin{figure}[ht!]
\begin{center}
\includegraphics[scale=0.4]{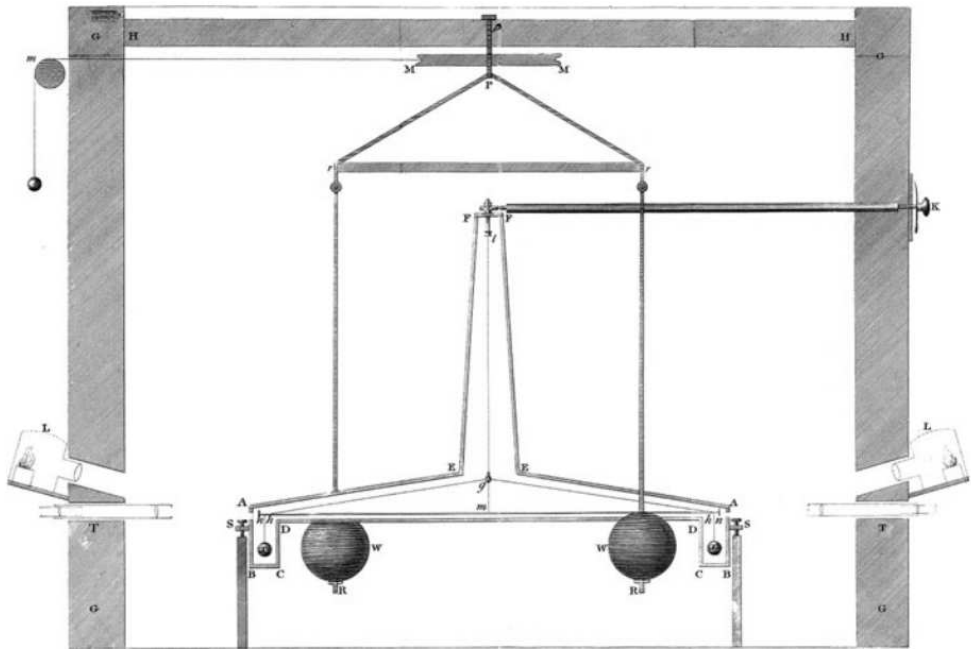}\\
\caption{El apparatus experimental de Cavendish, (sección vertical longitudinal). Tomada de Cavendish (\cite{1798}, p. 526).}\label{Ca1}
\end{center}
\end{figure}

\begin{figure}[ht!]
\begin{center}
\includegraphics[scale=0.4]{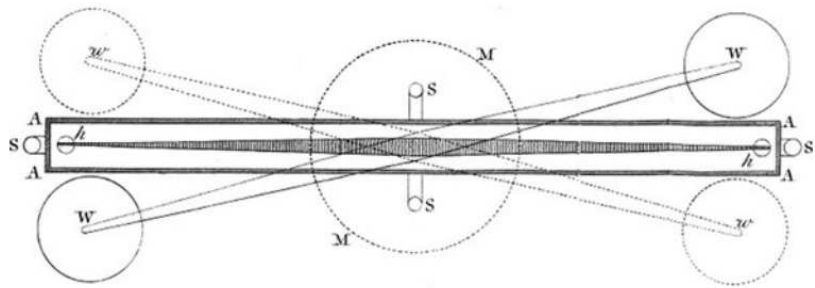}\\
\caption{Apparatus experimental de Cavendish (visto desde arriba). Tomado de Cavendish (\cite{1798}, p. 527).}\label{Ca2}
\end{center}
\end{figure}

El apparatus tomado para ser empleado en el famoso experimento de Cavendish\index{Experimento de Cavendish} fue un refinamiento de una apparatus original controvertido de John Michell (1724-1793)\footnote{La cooperación de Michell y Cavendish es discutida en la reciente biogarfía de McCormmach sobre Michell, \textit{Weighing the World: The Reverend John Michell of Thornhill}, Springer-Verlag. 2012.}, quien <<no completó el apparatus en poco tiempo antes de su muerte, y quien no viviría para realiza algún experimento con este>>. Después de su muerte este caería en las manos del Jackonian Professor en Cambridge, Francis John Hyde Wollaston, a quien, Cavensih, escribe, <<no tiene ventajas para realizar experimentos con esto, a la menera que él tendría que desaer, y sería bueno que él me diera esto a mí>> (Cf. Cavendish \cite{1798}, p. 464). La (Fig. \ref{Ca1}) muestra una sección vertical longitudinal a través del apparatus y la habitación, GGHHGG, en que este fue colocado.
\newline
La descripción de Cavendish\index{Cavendish} para el sistema experimental puede ser hallado en: Henry Cavendish 1798. \textit{Experiment to determine the density of the earth}. Philosophical Transactions of the Royal Society of London 88:469-526. En orden de preservar contra algunos orígenes de error, la habitación, midiendo 10 píes de altura y muchos píes cruzados, permanecería cerrada a través del experimento y los efectos serían observados desde el exterior de la habitación por medio de telescopios (T) y lámparas (L), que fueron instaladas a ambos lados de la habitación y que puntean los píes de Rey colocados en el interior de la casa. En esta dirección, el origen más significativo de error, a saber la variación de la temperatura, tendría que ser preservada significativamente, de acuerdo a Cavendish\index{Cavendish}. Dos bolas de plomo $x$ y $x$, que tienen un diámetro de alrededor de 2 pulgadas (o alrededor de 5.08 cm), son suspendidas por dos cables $hx$ del brazo $ghmh$ que por sí misma es suspendida por el cable delgado $gl$ con una longitud de alrededor de 40 pulgadas (o 1.016 m). Dado el hecho de que el cable es lo suficientemente delgado, <<la fuerza por minuto, tal como la atracción del peso de las bolas de plomo a unas pocas pulgadas del diámetro, será suficiente para atraer la balanza sensiblemente hacia un lado>>. Cavendish (\cite{1798}, p. 470) calculó que la fuerza por la que las bolas son atraídas en proporción a sus pesos es como de 1 a 50.000.000. Para determinar la fuerza por la que las bolas y la balanza son atraídas contra la fuerza restaurativa\index{Fuerza!restaurativa} para el cable rotado, la balanza fue colocada en tal dirección para permitir que esta se moviera libremente como un <<péndulo horizontal>>. La balanza $ghmh$ midiendo 6 pies (o aproximadamente 1.83 m) consiste de una delgada varilla $hmh$ reforzada por un cable de plata $hgh$, que <<es hecho lo suficientemente fuerte para soportar las bolas, a través de toda la luz>>. Las dos bolas $x$ y $x$ son colocadas en las cajas de madera estrechas $ABCDCBAEFFE$ que son tomadas horizontalmente y que es soportada por postes firmemente fijados en la base para la que este es pegado por cuatro tornillos $(S)$.\footnote{En la (Fig. \ref{Ca1}), la sección vertical longitudinal del apparatus, únicamente dos de los cuatro tornillos son ilustrados. Todos los cuatro tornillos son ilustrados en la Fig.2. Cavendish\index{Cavendish} notó que la caja en que las bolas son movidas tienen poca profundida <<lo que hace que el efecto de la corriente de aire sea más sensible de lo que debería ser, y es un defecto que yo tendré que corregir en algún experimento futuro>>. Cf. Cavendish (\cite{1798}, p. 497).} La caja de madera sirve para proteger la balanza de las corrientes de aire. $FK$ representa una barra de madera, que, por medio de un tornillo sin fin, gira alrededor del soporte y para el que el cable delgado $gl$ es atado. Por medio de $FK$ Cavendish tendría que manipular la posición para la balanza $ghmh$ del exterior de la misma obteniendo la posición requerida sin algún daño de contacto en el lado de la caja. El cable $gl$ es atado a su soporte en la cima y en el centro de la balanza y en la parte inferior por clips de hierro en el que este es reforzado por tornillos. Dos pesos conducidos $W$ y $W$ son suspendidos de las varillas de cobre $P_r$ y $rR$ y la barra de madera $rr$, que está entre las varillas. Esta pieza fue agregada al centro del broche $Pp$ que fue agregado al cieloraso $HH$ de la habitación y colocado bajo el centro del apparatus. Para $Pp$ la polea, $MM$, alrededor en que la cuerda $Mm$ fue atada así que uno puede alterar la posición de los pesos $W$ y $W$ desde el exterior. La (Fig. \ref{Ca2}) detalla el punto de vista por encima del instrumento.
\newline
Cuando los pesos $W$ y $W$ están en la primera posición~---indicada por líneas continuas~---ellas conspiran en la atracción de la balanza en la dirección $hW$; cuando los pesos están en la segunda posición~---indicada por líneas punteadas~---ellas atraen la balanza en la dirección contraria de $hW$. Como en la segunda posición la balanza fue atraída hacia fuera en tal dirección como para hacer el punto índice para un número superior sobre los resbalones de marfíl, Cavendish\index{Cavendish} consideró esto como la <<posición positiva para los pesos>>. Los pesos $W$ y $W$ fueron además prevenidos de golpear el instrumento por piezas de madera, atados al muro de la habitación, que detiene los pesos tan próntamente como ellos viene en el interior de 5 para una pulgada (o 0.508 cm) para el caso. Cavendish\index{Cavendish} encontró que <<los pesos pueden golpear contra ellos con considerable fuerza, sin una sensibilidad temblorosa en el instrumento>>. Además, <<[en] orden de determinar la situación de la balanza>>, trozos de marfil, que fueron divididos de a una veintena de una pulgada (o 1.27 mm), fueron colocados en el interior de la caja, tan cercanos para tocar el final de la balanza como podría ser posiblemente hecho sin tocarlos a ellos. Para los trozos originales sobre cada lado un nonius (o escala de Vernier) fue agragado, que en este fue dividido en el interior de cinco partes así que la posición para la balanza tendría que ser medida para 1 a 100 de una pulgada (i.e., para 0.254mm). Una vez la balanza es tomada en reposo y su posición fue observada, Cavendish (\cite{1798}, p. 474) movió los pesos $W$ y $W$ cercanos a las bolas $x$ y $x$ así que <<la balanza bien no únicamente es atraída hacia un lado por lo tanto, excepto que esto es adecuadamente hecho para vibrar, y sus vibraciones adecuadamente continuan por un buen rato>>.\\ Los intentos para determinar la dendidad de la Tierra fueron realizados antes que Cavendish por el experimento con la barra de torsión. Un método bien elaborado consistió en la medida de la deflexión de una plomada en la vecindad de una gran montaña. Este fue el método empleado por Nevil Maskelyne (1732-1811) en el famoso experimento en el monte Schiehallion en Escocia.\footnote{Cf. Maskelyne \cite{1775} para una discusión.} En un breve pero agudo artículo escrito por Jacob (1813-1862) éste señaló que <<el experimento de Cavendish\index{Experimento de Cavendish} es uno que puede ser confiado como dando una buena aproximación para la verdad, en el interior de los límites del error (cuando es conducido con adecuada precaución)>>. Cf. Jacob (\cite{Jacob1857}, p. 295).
\section{Medidas y su computación}
Después de haber proporcionado la descripción del sistema experimental, Cavendish\index{Cavendish} explicó como él hizo para determinar el punto de reposo de una vibración y el tiempo de vibración. Para establecer el punto de reposo, fue necesario <<observar los puntos extremos de las vibraciones, y por lo tanto determinar el punto en que este tendría que estar en reposo como si su movimiento fuera destruído, o el punto de reposo, como Yo tendré que llamar a este>>. Cf. Cavendish (\cite{1798}, p. 474). Para hacer esto, Cavendish, observó tres puntos extremos sucesivos de una vibración y tomó el medio entre el primero y el tercero de aquellos extremos, como el punto extremo de la vibración en una dirección, sobre el otro lado y, sobre el otro, el tomó el medio del punto extremo de la vibración y el segundo extremo como el punto de reposo, <<como las vibraciones son continuamente disminuidas>>, él observó, <<es evidente, que el medio entre dos puntos extremos no produzca el punto verdadero de reposo>>. Él entonces determinó el tiempo de vibración por obervar los dos puntos extremos de una vibración y los tiempos en que la balanza arribó a las dos divisiones entre los extremos. De lo anterior, él computó el punto medio para la vibración y, por proporción, el tiempo en el que la balanza arriba a este punto medio. Después de un número de vibraciones él repitió este procedimiento y dividió el intervalo de tiempo, entre el arribo de la balanza y los dos puntos medios, por el número de vibraciones, que da el tiempo de una vibración.\footnote{Cavendish (\cite{1798}, p. 478) nota que el error en el resultado es mucho menor, cuando las fuerzas requeridas para atraer la balanza hacia un lado fue deducida de experimentos hechos para cada experimento, cuando esto es tomado de experimentos previos.} <<Para juzgar la propiedad de este método>>, uno necesita considerar <<en que manera la vibración es afectada por la resistencia del aire, y por el movimiento del punto de reposo>>. Cf. Cavendish, (\cite{1798}, p. 476). Cavendish\index{Cavendish}, sin embargo, arguyó que en ambos casos el efecto es adecuadamente inconsiderable. Primero, <<como el tiempo de venida para el punto medio es anterior a la medida de la mitad para la vibración, ambos en la primera y última vibración, y en general así a su cercanía, el error producido de esta causa necesita ser inconsiderable>>. Segundo, así como el punto medio de reposo puede ser considerado como un movimiento uniforme, el tiempo de dos vibraciones sucesivas <<adecuadamente es muy poco alterado; y, entonces, el tiempo de movimiento del punto medio de una vibración para el punto medio del siguiente, también será muy poco alterado>>. Es relevante notar que Cavendish fue un meticuloso observador quien fue muy reconocido por la calibración de instrumentos científicos~---en efecto, él fue muy activo en los tiempos cuando los instrumentos de la Royal Society estaban siendo calibrados~---y <<en su trabajo experimental él mostró una profunda comprensión de la teoría de los errores>>.\\ El experimento de Cavendish\index{Experimento de Cavendish} es en efecto una concatenación de 17 experimentos relativos. Las determinaciones específicas para los movimientos de la balanza y los tiempos de vibración en cada uno de aquellos adecuadamente no es discutido: el autor tuvo que restringirse asi mismo a una discusión para los resultados obtenidos, que Cavendish resumió sobre la p. 520 de su artículo, y sus compuatciones. La tercera y quinta columna contienen las distancias atrevesadas por la balanza y los tiempos de vibración que fueron hallados en los 17 experimentos precedentes (cf. Fig. \ref{Ca3}). La segunda columna muestra las posiciones de partida para la balanza y las direcciones en que esta fue movida.\\ En los tres primeros experimentos, Cavendish (\cite{1798}, p. 478) empleó un cable de plata cobrizo, que, como él próntamente halló, no fue suficiente así que <<la atracción para el peso de atracción de la bola es mucho más aislado, así ellos tocan los lados de la caja>>. Sin embargo, él decidió hacer algunos experimentos con esto. En orden de garantizar que las vibraciones no fueran producidas por el magnetismo, él cambió los trazos de hierro, por lo que él permitió que los pesos fueran suspendidos, por unos de cobre, y como resultado de esto resultó que <<se observará que se tienen algunos efectos del mismo tipo, pero más irregulares, así que yo atribuyo esto a alguna causa accidental, y entonces en desacuerdo sobre los pesos permitidos, y procediendo con los experimentos>>. Cf. Cavendish\index{Cavendish} (\cite{1798}, p. 479). Además, Cavendish observó que:\\
<<si un cable es rotado únicamente un poco más de lo que admite su elasticidad, entonces, en lugar de nivelarse, como esto es llamado, o acquiriendo o permaneciendo rotado totalmente de una vez, esto es gradualmente, y, cuando este es dejado en libertad, este gradualmente pierde parte del conjunto que es acquirido; así que si, en este experimento, el cable, por haber preservado la rotación por 2 o 3 horas, tuvo gradualmente que producir esta presión, o tuvo que iniciar a tomarla, y este tendría gradualmente que restaurarse por si mismo, cuando se deja en libertad, el punto de reposo tendría gradualmente que moverse hacia atrás; pero piénsese que el experimento fue repetido dos veces, y yo no tuve que percibir algún tal efecto>>.\footnote{Cf. Cavendish, (\cite{1798}, p. 485).}
\section{?`Qué y cómo midió Cavendish?}
Cavendish trabajó en el interior de un marco matemático (similar a Newton\index{Newton}) basado sobre proporciones, donde la constante de gravitación universal puede únicamente ser concebida en el interior de un marco matemático de ecuaciones y medidas absolutas. Las Prop. VII y VIII del libro III del \textit{Principia}\index{Principia} conjúntamente establecen la ley de gravitación universal\index{Ley de gravitación universal}, pero, jústamente todas las proposiciones en el \textit{Principia} \textit{a calce ad capitum} (de pies a cabeza), sobre esta ley fueron expresadas verbalmente en términos de proposiciones:
\begin{propo}[\textbf{Prop. 7 del Principia}] La gravedad\index{Gravedad} existe en todos los cuerpos universalmente y es proporcional a la cantidad de materia en cada uno\footnote{Cf. Newton, (\cite{1999}, p. 810.)}.$\hspace{2.5cm}\blacksquare$
\end{propo}
\par
\begin{propo}[\textbf{Prop. 8 del Principia}]
Si dos globos gravitan uno hacia el otro y su materia es homogenea sobre todos los lados en las regiones que están a distancias iguales de sus centros, entonces el peso de cada uno de los globos hacia el otro bien será inversamente como el cuadrado de la distancia entre los centros\footnote{Ibid. p. 811}.$\hspace{2.5cm}\blacksquare$
\end{propo}
\par
En otras palabras, en el \textit{Principia}\index{Principia} de Newton\index{Newton} no hay indicios en los que pueda ser hallada la celebérrima ecuación $F=\dfrac{GMm}{r^2}$. En este contexto, el rol de $G$, i.e., $6.674\times 10^{-11}N\left(\dfrac{m^2}{Kg^2}\right)$, en la fórmula de la gravitación universal tendrá que ser explicada.\\ De la definición de densidad media para la Tierra sabemos que la densidad de la Tierra, $\rho(e)=\dfrac{m(e)}{V(e)}$, donde $m(e)$ es la masa de la Tierra y $V(e)$ su volumen. Además, de la definición de densidad y $F=\dfrac{Gm(e)m'}{R{(e)}^2}=gm'$, donde $R(e)$ es el radio de la Tierra y $m'$ es la masa de un cuerpo cercano a la superficie de la Tierra, de esto se sigue que
\begin{equation}\label{C1}
\rho(e)=\dfrac{gR{(e)}^2}{GV(e)}
\end{equation}
Si nosotros, además, aproximamos la figura de la Tierra a una esfera, se sigue que el volumen, $V(e)$, es igual a
\begin{equation}\label{C2}
\dfrac{4}{3}\pi R{(e)}^3
\end{equation}
Cuando nosotros completamos en esta determinación el volumen de la Tierra, en \eqref{C1}, nosotros obtenemos: $$\rho(e)=\dfrac{gR{(e)}^2}{G\dfrac{4}{3}\pi R{(e)}^3}=\dfrac{g}{G\dfrac{4}{3}\pi R(e)}$$ Hay que recordar que la primera medida para la constante de gravitación universal, $G$, únicamente apareció hacia finales del siglo XIX, y fue imposible para Cavendish\index{Cavendish} realmente calcular $G$ a partir de la densidad media de la Tierra.
\section{Derivando la densidad media de la Tierra por un razonamiento proporcional}
Para obtener una comprensión total del significado del experimento de Canvendish, es requerido que nosotros tengamos en cuenta el tratamiento origial de Cavendish, que estaba basado sobre proporciones, lo más cercano posible. Tenemos que comprender que Cavendish necesitó probar: una relación matemática en que la densidad media de la Tierra dada en términos del movimiento de la balanza y el tiempo de vibración.\\ Estos son los tres pasos relevantes en el argumento de Cavendish.
\begin{enumerate}
\item[\textbf{Paso 1:}] Resolver la proporción para la fuerza\index{Fuerza} que necesita ser aplicada a cada bola para atraer la balanza hacia un lado por una escala de
división para la fuerza de gravedad\index{Gravedad} sobre cada bola en términos del período de vibración de la balanza.
\end{enumerate}
\par
Primero que todo, Cavendish\index{Cavendish} determinó la fuerza requerida para atraer la balanza hacia un lado, que es determinada por el tiempo de una vibración. Él trató el movimiento de la balanza como un \textit{péndulo horizontal} que él comparó con el movimiento de un péndulo (vertical) regular. Dada la similaridad teórica entre ellos, Cavendish pudo transferir ciertas proporciones, que son fijas para un péndulo vertical, al péndulo horizontal a la mano. Porque la distancia entre los centros para las dos bolas, $x$ y $w$, es 73.3 in, la distancia de cada una al centro de movimiento es 36.65 in. Además, la longitud de la vibración de un péndulo en segundos <<en esta región>> es 39.14 in.\\ Entonces,\\
<<si la rigidez del cable por el que la balanza es suspendida en tal, la fuerza que necesita ser aplicada para cada bola, en orden de atraer la balanza hacia un lado por el ángulo $\widehat{A}$, es al peso de las bolas como el arco de $\stackrel{\frown}{A}$ es al radio,\footnote{Lo que Cavendish está estableciendo aquí es equivalente a decir que la fuerza restaurativa\index{Fuerza!restaurativa} del movimiento del péndulo $(F_r)$ a la vertical a través de un ángulo $\widehat{A}$ es al peso de las bolas por el $\sin(A)$.} la balanza adecuadamente vibra en el mismo tiempo como un péndulo cuya longitud es 36.65 in, esto es, en $\sqrt{\tfrac{36.65}{39.14}}\text{seg}$,\footnote{Como en este caso $\tfrac{x_1}{x_2}$ es proporcional a $\tfrac{t_1^1}{t_2^2}$ se sigue que $\sqrt{\tfrac{x_1}{x_2}}$ es proporcional a $\tfrac{t_1}{t_2}$. Si $x_1$ es 36.65 in, $x_2$ es 39.14 in y $t_2^2$ es 1, se sigue que $t_1$ es proporcional a $\sqrt{\tfrac{x_1}{x_2}}$ o, de lo que se obtiene, proporcional a $\sqrt{\tfrac{36.65}{39.14}}$.} y entonces, si la rigidez del cable es tal que para hacer que este vibre en $N\text{seg}$, la fuerza que necesita ser aplicada a cada bola, en orden de atraer esta hacia un lado por un ángulo $\widehat{A}$, es al peso de la bola como el arco de $\stackrel{\frown}{A}\times\tfrac{1}{N^2}\times\tfrac{36.65}{39.14}$ es al radio>>.\footnote{Cf. Cavendish, (\cite{1798}, p. 509.)}\\ Aquí Cavendish\index{Cavendish} señala que la fuerza ejercida sobre las bolas $(F_e)$ a lo largo de un péndulo es a la fuerza restauradora\index{Fuerza!restaurativa} $(F_r)$ como $\tfrac{T^2}{N^2}$. Porque la fuerza restauradora es, además, proporcional al peso de la bola $(W_b)$ por el arco de $\stackrel{\frown}{A}$, se sigue que:
$$\tfrac{F_e}{W_b}\squaredots\stackrel{\frown}{A}\times\tfrac{T^2}{N^2}\left(=\stackrel{\frown}{A}\times\tfrac{36.65}{39.14}\times\tfrac{1}{N^2}\right)$$
Como la escala de marfil al final de la balanza es de 38.3 in, de distancia al centro de movimiento y cada división es de $1/20$ de pulgada al centro de movimiento, este subtienda un ángulo al centro cuyo arco es $\tfrac{1}{766}$, i.e., $\tfrac{38.3 in}{0.05 in}$.\\ Por consiguiente, nosotros obtenemos que:
\begin{align}\label{C3}
&\dfrac{\footnotesize{\text{La fuerza que necesita ser aplicada a cada bola para atraerla hacia un lado por una división}}}{\footnotesize{\text{El peso de la bola}}}:\tfrac{1\times 36.65}{766N^2\times 39.14},\notag\\
\shortintertext{o,}
&\dfrac{\footnotesize{\text{La fuerza que necesita ser aplicada a cada bola para atraerla hacia un lado por una división}}}{\footnotesize{\text{El peso de la bola}}}\squaredots\tfrac{1}{818N^2}
\end{align}
Por depender sobre las propiedades matemáticas del péndulo, Cavendish\index{Cavendish} estableció una proporción que involucra el tiempo períodico.
\begin{enumerate}
\item[\textbf{Paso 2:}] Resolver la proporción para la fuerza de atracción\index{Fuerza!de atracci\'{o}n} del peso sobre su bola correspondiente para la fuerza de atracción de la Tierra sobre la bola en términos de la densidad de la Tierra relativa a la densidad del agua.
\end{enumerate}
\par
Segundo, es requerido encontrar <<la proporción con que la atracción del peso soporta acto seguido a la Tierra, suponiendo que la bola es colocada en la mitad de la caja, esto es, no debe estar más cercana que de un lado a otro>>. Cada uno de los pesos pesa 2.439.000 granos a \textit{grosso modo} 158 Kg,\footnote{Si nosotros asumimos que 1 grano es igual a 64.79891 mg, entonces, el peso $W,W_w$, corresponde a 158.044.541 mg o a 158Kg aproximadamente.} que es igual al peso de un píe esférico de agua 10.64, i.e., igual al peso de 10.64 veces el volumen de una esfera de agua con un diámetro de 1 píe.\footnote{De acuerdo a Cavendish, $W_w$ es igual a 10.64 veces el peso de una esfera de agua con un diámetro de 1 píe $(W_{\text{sph{1 píe}}}:W_w=158Kg=10.64\times W_{\text{sph(1 píe)}})$. Como $W_{\text{sph(1 píe)}}=V_{\text{sph(1 píe)}}\times\rho_{H_2O}$, se requiere que el valor de Cavendish introducido para la densidad del agua, $\rho_{H_2O}$, y el volumen de 1 pie esférico de agua, $V_{\text{sph(1 píe)}}$. Dado que el diámetro de 1 píe esférico de agua es 1 píe y píe es igual a 30.48cm, su volumen puede ser calculado, como sigue: $V_{\text{sph(1 píe)}}=\tfrac{4}{3}\times{(15.14cm)}^3=1.488\times 10cm^3=0.04183m^3$. En su paper, Cavendish no proporciona explícitamente un valor para $\rho_{H_2O}$. Sin embargo, el valor que él proporciona puede ser deducido. Dado que nosotros conocemos que: $$158Kg=10.64\times W_{\text{sph(1 píe)}}=10.64\times V_{\text{sph(1 píe)}}\times\rho_{H_2O}=10.64\times0.01483m^3\times\rho_{H_2O},$$ se sigue que: $\tfrac{158Kg}{10.64\times0.01483m^3=100\times 10\tfrac{Kg}{m^3}}$, que coinciden con el valor actual para la densidad del agua $\rho_{H_2O}$. Con esta información a la mano, se puede dar crédito a la afirmación de Cavendish: $W_w=158Kg=10.64\times W_{\text{sph(1 píe)}}$. Lo que nosotros necesitamos básicamente es una solución para el radio de $W_w$ para $W_{\text{sph(1 píe)}}$. Dado que $W_w=158Kg$ y $W_{\text{sph(1 píe)}}=0.01483m^3\times100\times10\tfrac{Kg}{m^3}$, se sigue que la proporción de $W_w$ para $W_{\text{sph(1 píe)}}$ es 16.7, que se aproxima al valor de Cavendish, 10.64. Cf. el excelente paper de Ducheyne (\cite{Ducheyne2011}, n. 24).} El radio de 1 píe esférico de agua es 6 in, como 1 píe es igual a 12 in. Entonces, Cavendish\index{Cavendish} continúo, la atracción de un peso sobre una partícula colocada en el centro de una bola a 8.85 in del centro de tal peso (denotado como $F_{W\rightarrow b}^{8.85 in}$) es para la atracción de 1 píe esférico de agua sobre una partícula colocada sobre esta superficie (denotado como $F_{sph(\text{1 píe})\rightarrow b}^{\text{sup}}$) como $10.64\times 0.9779\times{\left(\tfrac{6}{8.85}\right)}^2$ a 1, o:
$$\dfrac{F_{W\rightarrow b}^{8.85in}}{F_{sph(\text{1 píe})\rightarrow b}^{\text{sup}}}\squaredots\dfrac{10.64\times 0.9779\times{\left(\dfrac{6}{8.85}\right)}^2}{1}$$
Además, el diámetro medio para la Tierra es 41.800.000 píes y, entonces, si la densidad media de la Tierra es a la del agua como $D$ es a 1,
\begin{align}\label{C4}
\dfrac{\footnotesize{\text{La atracción del peso sobre la bola}}}{\footnotesize{\text{La atracción de la Tierra sobre la misma bola}}}\squaredots\dfrac{10.64\times 0.9779\times{\left(\dfrac{6}{8.85}\right)}^2}{41.800.000 D}\notag\\
\shortintertext{o,}
\dfrac{\footnotesize{\text{La atracción del peso sobre la bola}}}{\footnotesize{\text{La atracción de la Tierra sobre la misma bola}}}\squaredots\dfrac{1}{8.739.000 D}
\end{align}
Durante la derivación de \eqref{C4}, Cavendish\index{Cavendish} omitió algunos pasos cruciales. Aunque Cavendish no hace este punto explícito, esta conclusión estuvo basada sobre la ley de la gravitación universal de Newton. Esto fue de importancia absoluta, porque esta aplicación involucra la masa de la Tierra, que es igual a la densidad de la Tierra por el volumen de la misma. En otras palabras, esta está sobre la base de la ley de la gravitación universal y la definición de masa que la densidad (media) de la Tierra entra en el razonamiento matemático de Cavendish. Permítase que $F_{W\rightarrow b}$ sea el peso $W$ sobre la bola $x$ y $F_{E\rightarrow b}$ el peso de la Tierra sobre la bola $x$, y permítase que $d$ se refiera al diámetro, $\rho$ a la densidad, $m$ a la masa, y $r$ al radio. Dada la teoría de la gravitación universal, se sigue que:
\begin{align*}
&\dfrac{F_{W\rightarrow b}}{F_{E\rightarrow b}}\squaredots 10.64\times 0.9779\times{\left(\dfrac{6}{8.85}\right)}^2\times\dfrac{m\left(sph(\text{1 píe})\right)}{r{\left(sph(\text{1 píe})\right)}^2}\times\dfrac{rE^2}{mE},\\
\shortintertext{así que:}
&\dfrac{F_{W\rightarrow b}}{F_{E\rightarrow b}}\squaredots 10.64\times 0.9779\times{\left(\dfrac{6}{8.85}\right)}^2\times\rho_{H_2O}\times\dfrac{d{\left(sph(\text{1 píe})\right)}^3}{d{\left(sph(\text{1 píe})\right)}^2}\times\dfrac{dE^2}{\rho_E\times dE^2}\\
\shortintertext{Ahora, dado que $d\left(sph(\text{1 píe})\right)$ convenientemente igual a 1 in,}
&\dfrac{F_{W\rightarrow b}}{F_{E\rightarrow b}}\squaredots 10.64\times 0.9779\times{\left(\dfrac{6}{8.85}\right)}^2\times\dfrac{1}{\rho_{(E/H_2O)}}\times\dfrac{1}{dE}\\
\shortintertext{Entonces,}
&\dfrac{F_{W\rightarrow b}}{F_{E\rightarrow b}}\squaredots 10.64\times 0.9779\times\times{\left(\dfrac{6}{8.85}\right)}^2\times\dfrac{1}{D}\times\dfrac{1}{41.800.000},\\
\shortintertext{y, finalmente:}
&\dfrac{F_{W\rightarrow b}}{F_{E\rightarrow b}}\left(=\dfrac{\footnotesize{\text{La atracción para el peso sobre la bola}}}{\footnotesize{\text{La atracción para la Tierra sobre la masa de la bola}}}\right)\squaredots\dfrac{1}{8.739.000D}
\end{align*}
que es por tanto lo que Cavendish\index{Cavendish} afirmó en \eqref{C4}.
\begin{enumerate}
\item[\textbf{Paso 3:}] Combinando los pasos (1) y (2).
\end{enumerate}
\par
\begin{figure}[ht!]
\begin{center}
\includegraphics[scale=0.4]{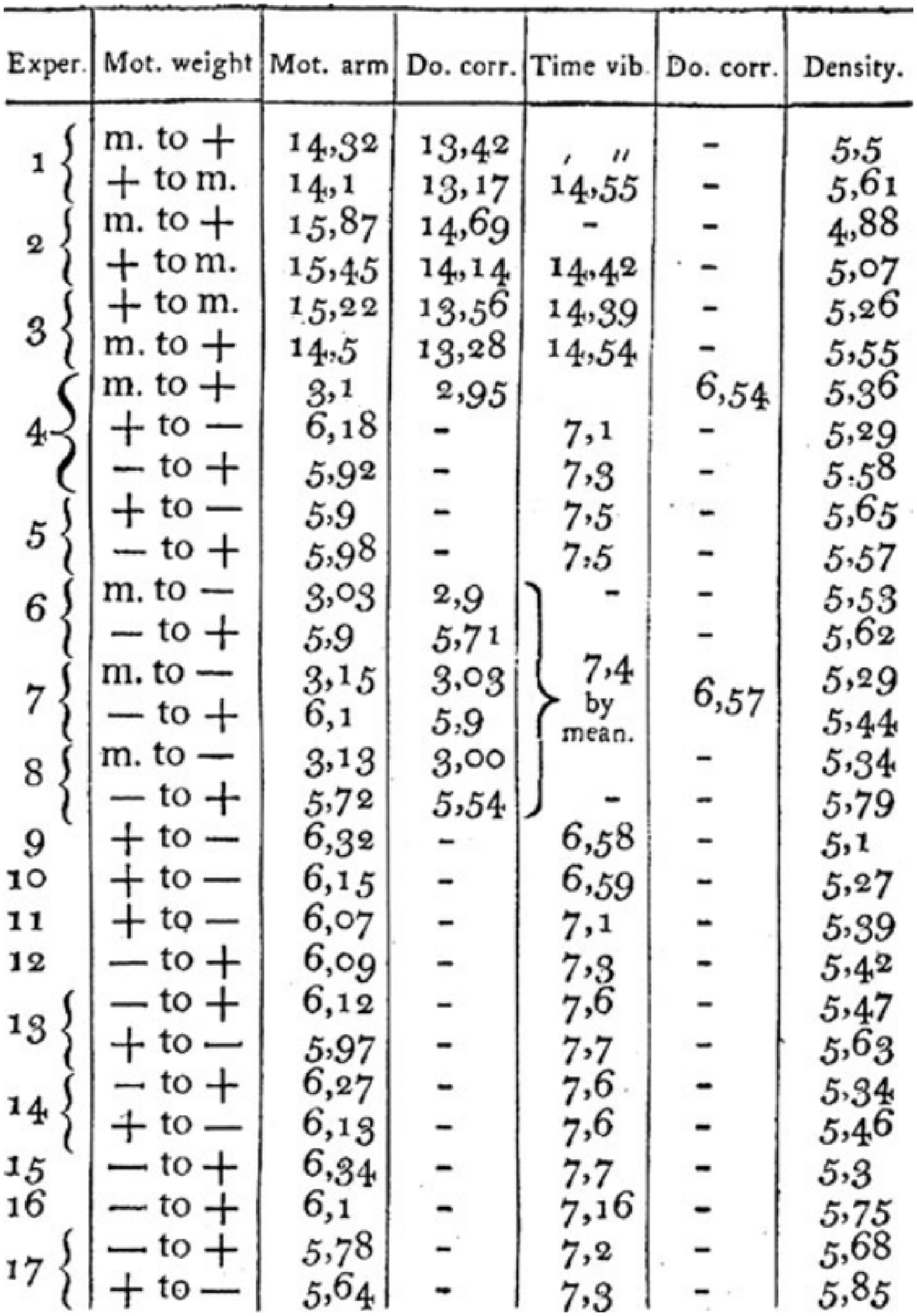}\\
\caption{Resumen de las medidas de Cavendish. Tomado de Cavendish (\cite{1798}, p. 520).}\label{Ca3}
\end{center}
\end{figure}

Previamente, nosotros tuvimos que establecer que la fuerza requerida para atraer la balanza a través de una división de 2.54 in es al peso de la bola como $1-818 N^2\,\textbf{Paso 1}$. Por dividir el \textbf{Paso 1} y el \textbf{Paso 2}, nosotros establecemos que:
\begin{align*}
&\cfrac{\footnotesize{\text{La fuerza que necesita ser aplicada a cada bola para ser atraía hacia un lado por 1 división}}}{\cfrac{\footnotesize{\text{El peso de la bola}}}{\cfrac{\footnotesize{\text{La atracción del peso sobre la bola}}}{\footnotesize{\text{La atracción de la Tierra sobre la misma bola}}}}}\squaredots\cfrac{1}{\cfrac{818N^2}{\cfrac{1}{8.739.000D}}},\\[.3cm]
&\tfrac{\footnotesize{\text{La atracción del peso sobre la bola}}}{\footnotesize{\text{La fuerza que necesita ser apliacada a cada bola para ser atraída hacia un lado por 1 división}}}\squaredots\tfrac{N^2}{10.683D},
\end{align*}
de lo que se sigue que <<la atracción [para el peso] bien atraerá la balanza por fuera de su posición natural por [...] $\tfrac{N^2}{10.683D}$ divisiones>>. Entonces, <<si sobre el movimiento de los pesos de la mitad del camino para una porción cercana de la balanza es hallado un movimiento $B$ divisiones, o si este se mueve $2B$ divisiones sobre el movimiento de los pesos de la balanza para una posición cercana al otro>>, la densidad media de la Tierra relativa a la densidad del agua $D$ es dada por $\tfrac{N^2}{10.683B}$, donde $B$ es el número de divisiones en centésimos de una pulgada y $N$ es el período observado en segundos (cf. Cavendish, \cite{1798}, p. 511). Por agregar factores de corrección (1) y (4), que adecuadamente serán discutidos en la nota a píe acompañante, para la fórmula anterior, Cavendish\index{Cavendish} corrigió $\tfrac{N^2}{10.683B}$ por $\tfrac{N^2}{10.844B}$ y por esta expresión él arribó a los valores en la columna 7 en la tabla sobre la (Fig. \ref{Ca3}).\footnote{Permítase, por instancia, computar el valor de $D$ para la entrada 1, <<\text{$+$ to m}>>. El tiempo es igual a $14'55''$ o 895seg y el movimiento para la balanza expresado en veinteavos de 1 pulgada es igual a 13.17. En efecto, $D$ es igual a $\tfrac{895^2}{13.17\times10.844}\approx 5.61$, que concuerda con el segundo resultado de Cavendish.}
\section{Determinaciones del siglo XIX para la constante gravitacional}
Como ya habíamos afirmado, la densidad media de la Tierra y la constante gravitacional son sencillamente derivables una de la otra: la primera es dada por $\tfrac{gR{(e)}^2}{GV(e)}$; y la última por $\tfrac{gR{(e)}^2}{\rho(e)V(e)}$. Además, ambas cantidades físicas fueron determinadas por la interacción gravitacional entre cuerpos calibrados en el laboratorio. Cerca de finales del siglo XIX, el foco de investigación tuvo que cambiar: la búsqueda de $G$ ahora ocupó un estado central, mientras la determinación de la densidad media de la Tierra fue concebida como un ejercicio directo sobre $G$ que tendría que ser determinado. Científicos activos en este campo se movieron más allá de presentar tales <<experimentos puramente locales>> como determinar la densidad media de la Tierra, entonces ellos determinaron la expresión exacta para la ley de la gravitación universal, y ellos ahora se consideraron a ellos mismos como <<trabajando para el universo>>.\footnote{Cf. Poynting, (\cite{1920}, p. 633).} En lo que sigue hablaremos un poco del brillante trabajo de John H. Poynting sobre la determinación de la constante gravitacional por medio de un ejemplo. También se mencionarán brevemente los resultados obtenidos por Charles V. Boys, Carl Braun y Franz Richarz y Otto Kriger-Menzel.\\ En orden de determinar $G$, Poynting suspendió dos masas iguales $A$ y $B$ (21.582.33 gr y 21.566.21 gr, respectivamente) de una balanza (cf. Poynting, \cite{1892}, p. 579). $A$ y $B$ estuvieron, además, colocadas en el interior de una caja de madera. A continuación, la masa $M$, pesando 153.407.269 gr, fue colocada por debajo de $A$. Una vez el cambio de posición de la viga ha sido observado, tendrá una longitud de 1.23329 m,\footnote{Ibib., p. 571} $M$ es rotada $180^\circ$ así que esta es colocada por debajo de $B$. Entonces, la posición de la viga es observada una vez más. Como $M$ cambia de lado entres $A$ y $B$, la atracción es tomada a lo lejos de $A$ y agregada a $B$. Para eliminar la atracción de $M$ sobre la viga y los cables suspendidos, $A$ y $B$ son colocados en posiciones igualmente altas $A'$ y $B'$, <<[para] la diferencia entre los dos incrementos de peso sobre la derecha, es devido solamente a la alteración de las posiciones de $A$ y $B$ relativas a $M$, la atracción sobre la viga permanece igual en cada una>>.\footnote{Ibid., p. 567}\\ En orden para compensar la inclinación del piso que surge cuando $M$ es movida, una masa adicional $m$, que es cercana a la mitad de la gran masa de $M$ (viz., 76.497.4 gr) fue instalada dos veces sobre el eje y sobre el lado opuesto de $M$.\footnote{Ibid., pp. 567-568, p. 579} Debido a la adición de $m$, la <<presión resultante fue ahora siempre a través del eje>> y no <<la inclinación del piso cuando la platarforma giratoria fue movida>> podría ser detectado. Ambas $M$ y $m$ fueron colacadas sobre la plataforma giratoria, que tendría que ser manipulada en la habitación bajo el sótano, en que el apparatus fue instalado. Una escala fue, además, fijada horizontalmente al final del telescopio por medio del que el jinete subsidiario se agregó al centro de la balanza y podría ser monitoreado, y por tanto la inclinación de la viga. Durante el experimento, las corrientes de aire y variaciones en la temperatura y la presión del aire fueron abolidas lo mejor posible. Sobre el proceso de la obtención de datos, Poynting estableció un valor para $G$ de $6.6984\times10^{-1}\tfrac{cm^3}{g\times s^2}$ (o, $6.6984\times10^{-11}\tfrac{m^3}{Kg\times s^2}$).\footnote{Ibid., p. 612} Una vez $G$ fue determinada, Poynting concluyó que la densidad media relativa para la densidad del agua es igual a 5.4934.\footnote{Ibid., p. 607}\\ Tres años después de la publicación del paper de Poynting, Boys publicó un nuevo método para determinar $G$. Sobre la computación de los datos obtenidos, Boys infirió que el valor para $G$ es igual a $6.6579\times10^{-8}\tfrac{cm^3}{g\times s^2}$ (o, $6.6579\times10^{-11}\tfrac{m^3}{Kg\times s^2}$).\footnote{Cf. Boys, (\cite{Boys1895}, p. 62).} De esto él obtuvo un valor de 5.5268 para la densidad media de la Tierra. Un año después de la publicación del paper de Boys, Braun publicó una nueva determinación para la constante gravitacional. Braun estableció un valor de $6.65816\pm0.00168\times10^{-8}\tfrac{cm^3}{g\times s^2}$ (o, $6.65816\pm0.00168\times10^{-11}\tfrac{m^3}{Kg\times s^2}$) para la constante gravitacional; correspondientemente, él arribó a un valor de $5.52700\pm0.0014$ para la dendidad media de la Tierra relativa a la densidad del agua.\footnote{Cf. Braun (\cite{1896}, p. 258).}\\ Finalmente, en 1898 Richarz y Krigar-Menzel publicaron los resultados que ellos obtuvieron al experimentar con otro artefacto. En tal paper conjunto, ellos concluyeron que la constante gravitacional es igual a $6.685\pm0.011\times10^{-8}\tfrac{cm^3}{g\times s^2}$ (o, $6.685\pm0.011\times10^{-11}\tfrac{m^3}{Kg\times s^2}$) y, sobre la base de esta determinación, ellos arribaron al valor de $5.505\pm0.009$ para la densidad media de la Tierra relativa a la densidad del agua.\footnote{Cf. Richarz and Krigar-Menzel  (\cite{1898}, p. 110)}\\ En aquellos experimentos de finales del siglo XIX, factores no-gravitacionales adicionales estuvieron fuera de cartelera (i.e., presión atmosférica y la humedad atmosférica) y variaciones en la temperatura y corrientes de aire estuvieron por fuera de cartelera más estríctamente. Por la eliminación de perturbaciones no-gravitacionales más satisfactoriamente, los físicos estarían más satisfactoriamente aislando las interacciones gravitacionales de los cuerpos en el laboratorio, sobre la base de que ellos podrían determinar la gran $G$. En concordancia, ellos establecerían un cubrimiento creciente e incrementarían la exactitud de las medidas para la gran $G$ y la densidad media de la Tierra.

\phantomsection
\nocite{*}
\bibliographystyle{abbrvnat}
\bibliography{Biblio}

\end{document}